\documentclass[journal]{IEEEtran}
\usepackage{makeidx}
\usepackage{amsfonts}
\usepackage{cite}
\usepackage{url}
\usepackage{epsfig,dsfont,amsmath,amssymb}
\usepackage{wrapfig,floatflt}
\usepackage{theorem}
\usepackage{graphics}
\usepackage{subfig}
\usepackage{stfloats}
\usepackage{multirow}
\usepackage{footnote}
\usepackage{hyperref}
\hypersetup{
colorlinks = false;
linkcolor = red;
citecolor = green;
}
\makesavenoteenv{tabular}
\graphicspath{{./images/}}

\def \bs {\boldsymbol}
\def \Pr {\mathbb{P}}

\def \tr {\mathrm{trace}}

\def \I {\mathrm{I}}
\def \R {\mathbb{R}}

\def \a {\bs{a}}
\def \sF {\mathcal{F}}
\def \X {\bs{X}}
\def \Q {\mathcal{Q}}

\def \y {\bs y}
\def \x {\bs x}

\def \hx {\hat{\x}}
\def \z {\bs z}
\def \w {\bs w}
\def \h {\bs h}
\def \H {\mathcal{H}}

\def \r {\bs r}
\def \u {\bs u}

\def \g {\bs g}

\def \sx {\mathcal{X}}

\def \zero {\bs 0}
\def \N {\mathcal N}
\def \E {\mathbb E}

\def \df {\stackrel{\mathrm{def}}{=}}

\def \supp {\mathrm{supp}}

\def \sup {\mathrm{sup}}
\def \mean {\mathrm{mean}}
\def \std {\mathrm{std}}
\def \nn {\nonumber}

\newtheorem{thm}{Theorem}

\newtheorem{defi}{Definition}

\title{Performance Analysis of Sparse Recovery Based on Constrained Minimal Singular Values}

\author{\IEEEauthorblockN{Gongguo Tang, \emph{Member, IEEE} and Arye Nehorai, \emph{Fellow, IEEE}}\thanks{This work was supported by the Department of Defense under the Air Force Office of Scientific Research MURI Grant FA9550-05-1-0443, ONR Grant N000140810849, and the National Science Foundation, Grant No.\ CCF-1014908.}\\
\thanks{Gongguo Tang and Arye Nehorai are with the Preston M. Green Department of Electrical and Systems Engineering, Washington University in St. Louis, St. Louis, MO 63130-1127, (email: nehorai@ese.wustl.edu)}
}
\begin{document}

\maketitle
\noindent
\begin{abstract}
The stability of sparse signal reconstruction is investigated in this paper. We design efficient algorithms to verify the sufficient condition for unique $\ell_1$ sparse recovery. One of our algorithm produces comparable results with the state-of-the-art technique and performs orders of magnitude faster. We show that the $\ell_1$-constrained minimal singular value ($\ell_1$-CMSV) of the measurement matrix determines, in a very concise manner, the recovery performance of $\ell_1$-based algorithms such as the Basis Pursuit, the Dantzig selector, and the LASSO estimator. Compared with performance analysis involving the Restricted Isometry Constant, the arguments in this paper are much less complicated and provide more intuition on the stability of sparse signal recovery. We show also that, with high probability, the subgaussian ensemble generates measurement matrices with $\ell_1$-CMSVs bounded away from zero, as long as the number of measurements is relatively large. To compute the $\ell_1$-CMSV and its lower bound, we design two algorithms based on the interior point algorithm and the semi-definite relaxation.
\end{abstract}

\vspace{-0.3cm}
\section{Introduction}
\label{sec:intro}
\noindent

Sparse signal reconstruction aims at recovering a sparse signal $\x \in \R^n$ from observations of the following model:
\begin{eqnarray}\label{intro:model}
  \y = A\x +\w,
\end{eqnarray}
where $A\in \R^{m\times n}$ is the measurement or sensing matrix, $\y$ is the measurement vector, and $\w \in \R^m$ is the noise vector. The sparsity level $k$ of $\x$ is defined as the number of non-zero components of $\x$. The measurement system is underdetermined because the number of measurements $m$ is much smaller than the signal dimension $n$. However, when the sparsity level $k$ is also small, it is possible to recover $\x$ from $\y$ in a stable manner. Reconstruction of a sparse signal from linear measurements appears in many signal processing branches, such as compressive sensing \cite{Candes2006Uncertainty, Donoho2006Compressed, Candes2005Decoding}, sparse linear regression \cite{Larsson2007Regression}, source localization \cite{Willsky2005Source, Model2006Signal}, sparse approximation, and signal denoising \cite{donoho1989uncertainty}. Model \eqref{intro:model} is applicable to many practical areas such as DNA microarrays \cite{Baraniuk2007DNA,Vikalo2007DNA,Parvaresh2008DNA,Vikalo2008DNA}, radar imaging \cite{Baraniuk2007Radar,Herman2008Radar,Herman2008Highradar},
cognitive radio \cite{Tian2007Cognitive}, and sensor arrays \cite{Willsky2005Source, Model2006Signal}, to name a few.

This paper is motivated by the following considerations.  When we are given a sensing or measurement system \eqref{intro:model}, we usually want to know the performance of the system before using it, or at least to know whether it works in the ideal setting. This involves deriving a verifiable sufficient condition for unique recovery and a computationally amenable performance measure. Furthermore, in signal processing and the applications mentioned in the previous paragraph, we usually have the freedom to design the sensing matrix; that is, we can choose the best from a collection of sensing matrices. For example, in radar imaging and sensor array applications, sensing matrix design is connected with waveform design and array configuration design. The $\ell_1$-CMSV of this paper has already been used to optimally design orthogonal frequency division multiplexing (OFDM) radar signals for detecting a moving target in the presence of multipath reflections \cite{Sen2010ofdm}. We hope the practitioners in similar fields will also find the results in this paper useful.

The contribution of the work is fourfold. First, we design two algorithms to verify the sufficient condition that a sparse signal can be uniquely recovered using $\ell_1$ minimization in the noise-free setting. By solving multiple linear programs efficiently, one of the algorithms produces comparable results with the state-of-the-art verification algorithms and performs orders of magnitude faster. Second, we derive concise bounds on the $\ell_2$ norm of estimation error for the Basis Pursuit, the Dantzig selector, and the LASSO estimator in terms of the $\ell_1-$CMSV. As the third contribution, we demonstrate that if the number of measurements $m$ is reasonably large, subgaussian random matrices have $\ell_1$-CMSVs bounded away from zero, with high probability. This implies that at least for subgaussian random matrices, the $\ell_1$-CMSV is as good as the restricted isometry constant. Last but not least, we develop algorithms to compute the $\ell_1$-CMSV for an arbitrary sensing matrix and compare their performance. These algorithms are by no means the most efficient ones. However, once we shift from an optimization problem with a discrete nature (\emph{e.g.}, the restricted isometry constant) to a continuous one, there are many optimization tools available and more efficient algorithms can be designed.

Many quantities and properties on the sensing matrix $A$ have been proposed to guarantee a stable or unique signal reconstruction, most notably the Restricted Isometry Constant (RIC) \cite{Candes2006Uncertainty, Candes2008RIP} and the Null Space Property (NSP) \cite{Cohen2009NSP}. The RIC provides a unified framework to deal with sparse signal recovery and has very nice geometrical explanations. However, it is known to be very difficulty to compute. Therefore, computable bounds on quantities involved in the RIC and the NSP are computed using, for example, semi-definite programming relaxation \cite{dAspremont2007sparsePCA, dAspermont2010Nullspace}, and linear programming relaxation \cite{Juditsky2010Verifiable}. To the best of the authors' knowledge, the algorithms of \cite{dAspermont2010Nullspace} and \cite{Juditsky2010Verifiable} in verifying the sufficient condition of unique $\ell_1$ recovery represent the state-of-the-art technique in this direction. In our paper, instead of computing a quantity ($\alpha_k$ in \cite{Juditsky2010Verifiable}) for various sparsity levels $k$ and checking if it is less than $1/2$, we directly seek the critical sparsity level below which unique recovery is guaranteed. We compare our verification algorithms with those in \cite{dAspermont2010Nullspace} and \cite{Juditsky2010Verifiable} using numerical simulations. One of our algorithms performs orders of magnitude faster, consumes much less memory, and produces comparable results.

The paper is organized as follows. In Section \ref{sec:model}, we present the measurement model, three convex relaxation algorithms, and the RIC. Section \ref{sec:bound} is devoted to deriving sufficient conditions for unique $\ell_1$ recovery and deriving bounds on the recovery errors of several convex relaxation algorithms. In Section \ref{sec:random}, we show that the majority of realizations of the subgaussian measurement ensemble have good $\ell_1$-CMSVs. In Section \ref{sec:computation}, we design algorithms to verify unique recovery, and compute the $\ell_1$-CMSV and its lower bound. We compare the algorithms' performance in Section \ref{sec:simulations}. Section \ref{sec:conclusions} summaries our conclusions.

\section{The Measurement Model, Reconstruction Algorithms, and Restricted Isometry Property}\label{sec:model}

\subsection{The Measurement Model}\label{subsec:measmodel}
The following measurement model is used throughout the paper. Suppose we have a sparse signal $\x \in \R^n$, \emph{i.e.}, $\x$ has only a few non-zero components. The sparsity level $k$ of $\x$ is defined as the number of non-zero elements of $\x$, or the $\ell_0$ ``norm" of $\x$: $k = \|\x\|_0$. We call a vector $k-$sparse if its sparsity level $\|\x\|_0 \leq k$. For ease of presentation, we restrict ourselves to exactly sparse signals in this paper and leave approximately sparse signals to future work.

We observe $\y \in \R^m$ through the following linear model:
\begin{eqnarray}\label{model}
  \y &=& A\x + \w,
\end{eqnarray}
where $A \in \R^{m\times n}$ is the sensing/measurement matrix and $\w \in \R^m$ is the noise/disturbance vector.

In this paper, we focus on three most renown recovery algorithms based on convex relexation: the Basis Pursuit (BP) \cite{Donoho1998Atomic}, the Dantzig Selector (DS) \cite{Candes2007Dantzig}, and the LASSO estimator \cite{Tibshirani1996Lasso}:
\begin{eqnarray}
\hskip -1cm &&\text{BP:}\min_{\z \in \R^n}\|\z\|_1 \text{\ \ s.t.\ } \|\y - A\z\|_2 \leq \epsilon\label{bp}\\
\hskip -1cm && \text{DS:} \min_{\z \in \R^n}\|\z\|_1 \text{\ \ s.t. \ } \|A^T(\y - A\z)\|_\infty \leq \lambda_n \sigma\label{ds}\\
\hskip -1cm &&\text{LASSO:} \min_{\z \in \R^n} \frac{1}{2}\|\y - A\z\|_2^2 + \lambda_n\sigma \|\z\|_1 \label{lasso}.
\end{eqnarray}
Here $\lambda_n$ is a turning parameter, and $\epsilon$ and $\sigma$ is a measure of the noise level. All these three optimization programs can be implemented efficiently using convex programming or even linear programming.

The performance of the BP, the DS and the LASSO, more specifically the error bounds on the solutions of these algorithms, usually involve the incoherence of the sensing matrix $A$. Many quantities are proposed to measure the incoherence of the sensing matrix, for example, the Restricted Isometry Constant (RIC) \cite{Candes2008RIP, Candes2005Decoding}, the Restricted Eigenvalue assumption \cite{bickel2009simultaneous}, and the Restricted Correlation assumption \cite{bickel2007discussion}, among others. However, these quantities are very difficult to compute. For example, the only known technique to exactly compute the RIC is test all its submatrices of certain size.

We will compare our CMSV based bounds with the RIC based bounds. For this purpose, we follow \cite{Candes2008RIP, Candes2005Decoding} to define the RIC as follows:
\begin{defi}
For each integer $k \in \{1,\ldots,n\}$, the restricted isometry constant (RIC) $\delta_k$ of a matrix $A \in \R^{m\times n}$ is defined as the smallest $\delta > 0$ such that
\begin{eqnarray}
1-\delta \leq \frac{\|A\x\|_2^2}{\|\x\|_2^2} \leq 1+\delta
\end{eqnarray}
holds for arbitrary non-zero $k-$sparse signal $\x$.
\end{defi}
The RIC has very clear geometrical meanings. A matrix $A$ with a small $\delta_k$ roughly means that $A$ is nearly an isometry when restricted onto all $k-$sparse vectors.


Now we cite some of the most renown performance results on the BP, the DS, and the LASSO, which are expressed in terms of the RIC. Assume $\x$ is a $k-$sparse signal and $\hx$ is its estimate given by any of the three algorithms; then we have the following:
\begin{enumerate}
  \item BP \cite{Candes2008RIP}: Suppose that $\delta_{2k} < \sqrt{2}-1$ and $\|\w\|_2 \leq \epsilon$. The solution to the BP \eqref{bp} satisfies
  \begin{eqnarray}\label{bp_rip_bd}
\|\hx-\x\|_2 \leq \frac{4\sqrt{1+\delta_{2k}}}{1-(1+\sqrt{2})\delta_{2k}}\cdot \epsilon.
\end{eqnarray}
  \item DS \cite{Candes2007Dantzig}: If the noise $\w$ satisfies $\|A^T\w\|_{\infty} < \lambda_n \sigma$, and $\delta_{2k} + \delta_{3k} < 1$. Then, the error signal obeys
      \begin{eqnarray}\label{ds_rip_bd}
        \|\hx-\x\|_2 \leq \frac{4\sqrt{k}}{1-\delta_{2k}-\delta_{3k}} \lambda_n \sigma.
      \end{eqnarray}
  \item LASSO \cite{meinshausen2009lassotype}: Consider the noise-free case. Under the condition of incoherence design with a sparsity multiplier sequence $e_n$, the error associated with the LASSO estimator $\hx$ is bounded for sufficiently large $n$ by
      \begin{eqnarray}\label{lasso_rip_bd}
        \|\hx - \x\|_2 \leq 17.5\cdot \lambda_n\sigma \cdot \frac{\sqrt{k_n}}{(\nu_{e_nk_n}^{\min})^2}.
      \end{eqnarray}
      Here, the sparsity level $k = k_n$ depends on $n$. Refer to \cite{meinshausen2009lassotype} for more information on incoherence design and multiplier sequence.
\end{enumerate}
We note that in these error bounds, the terms involving the RIC on the right hand sides are quite complicated. We will compare these results with our bounds in Section \ref{sec:bound}, which are much more concise and whose derivations are much less involved.

Although the RIC provides a measure quantifying the goodness of a sensing matrix, as mentioned earlier, its computation poses great challenge. The computation difficulty is compensated by the nice properties of RIC for a large class of random sensing matrices. We cite one general result below \cite{Baraniuk2008RIP}:
\begin{itemize}
  \item  Let $A \in \R^{m\times n}$ be a random matrix whose entries are \emph{i.i.d.} samples from any distribution that satisfies the concentration inequality for any $\x \in \R^n$ and $0 < \varepsilon <1$:
      \begin{eqnarray}\label{concentration}
        \Pr\left(\left|\|A\x\|_2^2-\|\x\|_2^2\right|\geq \varepsilon \|\x\|_2^2\right) \leq 2e^{-mc_0(\varepsilon)}.
      \end{eqnarray}
      Then, for any given $\delta \in (0,1)$, there exist constants $c_1, c_2 > 0$ depending only on $\delta$ such that $\delta_k \leq \delta$, with probability not less than $1-2e^{-c_2m}$, as long as
      \begin{eqnarray}\label{mk_old}
        m \geq c_1 k\log\frac{n}{k}.
      \end{eqnarray}
\end{itemize}
We remark that distributions satisfying the concentration inequality \eqref{concentration} include the Gaussian distribution and the Bernoulli distribution. For the $\ell_1$-CMSV, we will establish a theorem similar to the one above for the subgaussian ensemble with the same bound on $m$. The subgaussian ensemble in this paper includes the Gaussian ensemble, the Bernoulli ensemble, as well as the normalized volume measure on various convex symmetric bodies , for example, the unit balls of $\ell_p^n$ for $2 \leq p \leq \infty$ \cite{mendelson2007subgaussian}.

\section{Stability of Convex Relaxation based on the $\ell_1$-Constrained Minimal Singular Value}\label{sec:bound}
In this section, we present two approaches to verify the sufficient condition for the uniqueness of $\ell_1$-recovery. We also derive bounds on the reconstruction errors for the BP, the DS and the LASSO. Our bounds are given in terms of the $\ell_1$-CMSV rather than the RIC of matrix $A$ .

We first introduce a quantity that measures the sparsity, (or, more accurately, the density), of a given vector $\x$.
\begin{defi}
The $\ell_1$-sparsity level of a non-zero vector $\x \in \R^n$ is defined as
\begin{eqnarray}
  s(\x) = \frac{\|\x\|_1^2}{\|\x\|_2^2}.
\end{eqnarray}
\end{defi}

The scaling and permutation invariant $s(\x)$ is indeed a measure of sparsity. To see this, suppose $\|\x\|_0 = k$; then the Cauchy-Schwartz inequality implies that
\begin{eqnarray}\label{sp_bd}
  s(\x)  \leq k,
\end{eqnarray}
and we have equality if and only if the absolute values of all non-zero components of $\x$ are equal. Therefore, the more non-zero elements $\x$ has and the more evenly the magnitudes of these non-zero elements are distributed, the larger $s_p(\x)$. In particular, if $\x$ has exactly one non-zero element, then $s(\x) = 1$; if $\x$ has $n$ non-zero elements with the same magnitude, then $s(\x) = n$.

We use the $\ell_1-$sparsity level as a tool to relax the necessary and sufficient condition for exact $\ell_1$ recovery in the noise less setting \cite{zhang2005overunder, donoho2001uncertainty, donoho2004highdimensional}. In particular, Zhang showed in \cite{zhang2005overunder} that $\x$ with $\|\x\|_0 = k$ is the unique solution to BP with $\epsilon = 0$:
\begin{eqnarray}
  \min_{\z\in \R^n} \|\z\|_1 \text{\ s.t.\ } A\x = A\z
\end{eqnarray}
if and only if
\begin{eqnarray}\label{nullspaceproperty}
  \sum_{i\in S}|\z_i| < \sum_{i \notin S} |\z_i|
\end{eqnarray}
for any $\z$ such that $A\z = 0$ and any index set $S \subset \{1,\ldots,n\}$ of size at most $k$. We are interested in finding $k^*$, the maximal $k$ such that the necessary and sufficient condition \eqref{nullspaceproperty} is satisfied.

We note that a sufficient condition for exact $\ell_1$ recovery is
\begin{eqnarray}\label{l1suff}
  s(\z) &>& 4k
\end{eqnarray}
for any $\z \in \mathrm{Ker}(A) \df \{\z: A\z = 0\}$. This is because the negation of \eqref{nullspaceproperty}:
\begin{eqnarray*}
 &&\exists \z \in \mathrm{Ker}(A) \text{\ and \ } S \text{\ with size at most \ }k \\
 &&\text{\  such that\ } \sum_{i\in S}|\z_i| \geq \sum_{i \notin S} |\z_i|
\end{eqnarray*}
implies
\begin{eqnarray*}
  \|\z\|_1 &\leq & 2\sum_{i\in S}|\z_i|\\
  & \leq & 2\sqrt{k}\sqrt{\sum_{i\in S}|\z_i|^2}\\
  &\leq & 2\sqrt{k} \|\z\|_2.
\end{eqnarray*}
Therefore, the minimization of $s(\z)$ over $\mathrm{Ker}(A)$ yields a lower bound on $k^*$. Unfortunately, this optimization is very difficult. In section \ref{sec:computation}, we present a semidefinite relaxation algorithm to obtain a lower bound on $k^*$.

Another relaxation approach is to replace the $\ell_2$ norm in the definition of the $\ell_1$-sparsity level with the $\ell_\infty$ norm. Note that the negation of \eqref{nullspaceproperty} also implies that
\begin{eqnarray}
 \|\z\|_1 &\leq & 2\sum_{i\in S}|\z_i|\\
  & \leq & 2k \|\z\|_\infty.
\end{eqnarray}
Therefore, the following optimization problem
\begin{eqnarray}\label{eqn:maxfrac1inf}
  \min_{\z: A\z = 0} \frac{1}{2}\frac{\|\z\|_1}{\|\z\|_\infty}
\end{eqnarray}
finds a lower bound on the maximal $k$ such that \eqref{nullspaceproperty} is satisfied. In Section \ref{sec:computation}, we will present a polynomial time algorithm to solve \eqref{eqn:maxfrac1inf}. The algorithm solves $n$ linear programs and produces results comparable to the best known results in \cite{Juditsky2010Verifiable} within a much shorter time.

In the noisy setting, our derivation of the error bounds for the BP, the DS, and the LASSO relies heavily on the fact that the error vectors have small $\ell_1-$sparsity levels.\\

Now we are ready to define the $\ell_1-$constrained minimal singular value (CMSV):
\begin{defi}\label{def:l1cmsv}
For any $s \in [1,n]$ and matrix $A\in \R^{m\times n}$, define the $\ell_1$-constrained minimal singular value (abbreviated as $\ell_1$-CMSV) of $A$ by
\begin{eqnarray}
\rho_s(A) &\df& \min_{\x \neq 0,\ s(\x) \leq s } \frac{\|A\x\|_2}{\|\x\|_2}.
\end{eqnarray}
\end{defi}

Intuitively, the $\ell_1-$CMSV $\rho_s(A)$ measures the invertibility of the operator $A: \R^n \mapsto \R^m$ when restricted onto vectors with $\ell_1$-sparsity level not greater than $s$. \\

In the following theorem, we present our error bounds in terms of the $\ell_1-$CMSV, whose proof is given in Appendix \ref{sec:pf:errorbd}:

\begin{thm}\label{thm:errorbd}\vspace{-0.4cm}
Suppose the support of the true signal $\x$ is of size $k$.
\begin{enumerate}
  \item If the noise $\w$ is bounded; that is, $\|\w\|_2 \leq \epsilon$, then the solution $\hx$ to the BS \eqref{bp} obeys
\begin{eqnarray}\label{bd:bp}
  \|\hx - \x\|_2 &\leq& \frac{2\epsilon}{\rho_{4k}}.
\end{eqnarray}
  \item If the noise $\w$ in the DS \eqref{ds} satisfies $\|A^T\w\|_\infty \leq \lambda_n \sigma$, then the solution to \eqref{ds} obeys
\begin{eqnarray}\label{bd:ds}
\|\hx - \x\|_2 \leq \frac{4\sqrt{k}}{\rho_{4k}^2}\lambda_n \sigma.
\end{eqnarray}
  \item If the noise $\w$ in the LASSO \eqref{lasso} satisfies $\|A^T\w\|_\infty \leq \kappa\lambda_n\sigma$ for some $\kappa \in (0, 1)$, then the solution $\hx$ to the LASSO estimator \eqref{lasso} obeys
\begin{eqnarray}\label{bd:lasso}
\|\hx - \x\|_2 \leq \frac{1 + \kappa}{1-\kappa}\cdot \frac{2\sqrt{k}}{\rho_{\frac{4k}{(1-\kappa)^2}}^2}\lambda_n\sigma.
\end{eqnarray}
\end{enumerate}
\end{thm}

As shown in Appendix \ref{sec:pf:errorbd}, the procedure of establishing Theorem \ref{thm:errorbd} has two steps:
\begin{enumerate}
  \item For all three algorithms, show that the error vector $\h = \hx - \x$ is $\ell_1-$sparse: $s(x) \leq s$, where $s = 4k$ for the BP and the DS, and $s = 4k/(1-\kappa)^2$ for the LASSO. This automatically leads to a lower bound $\|A\h\|_2 \geq \rho_{s} \|\h\|_2$;
  \item Obtain an upper bound on $\|A\h\|_2$ or $\|A\h\|_2^2$ and invoke Definition \ref{def:l1cmsv} of the $\ell_1$-CMSV .
\end{enumerate}
The derivation is simpler than those employed for obtaining the RIC based bounds.

When the noise $\w \sim \N(0,\sigma^2\I_m)$, as shown by Cand\'es and Tao in \cite{Candes2007Dantzig}, with high probability, $\w$ satisfies the orthogonality condition
\begin{eqnarray}\label{orth}
  |\w^TA_j| &\leq& \lambda_n\sigma \ \ \ \ \text{for all}\ \ 1\leq j \leq n,
\end{eqnarray}
for $\lambda_n = \sqrt{2\log n}$. More specifically, defining the event
\begin{eqnarray}
E \df \{\|A^T\w\|_\infty \leq \lambda_n\sigma\},
\end{eqnarray}
we have
\begin{eqnarray}
  \Pr(E^c) &\leq& \frac{2n\cdot (2\pi)^{-1/2}e^{-\lambda_n^2/2}}{\lambda_n}.
\end{eqnarray}
Therefore, with $\lambda_n = \sqrt{2(1+t)\log n}$, we obtain
\begin{eqnarray}
  \Pr(E)\geq 1-\left(\sqrt{\pi(1+t)\log n}\cdot n^t\right)^{-1}.
\end{eqnarray}
As a consequence, the conditions on noise in Theorem \ref{thm:errorbd} for the DS and the LASSO hold with high probability.

Compared with the RIC bounds \eqref{bp_rip_bd}, \eqref{ds_rip_bd}, and \eqref{lasso_rip_bd}, our CMSV bounds \eqref{bd:bp}, \eqref{bd:ds}, and \eqref{bd:lasso} are more concise. Of course, if the CMSV $\rho_s(A)$ is not bounded away from zero, these concise bounds would not offer much. We will show in Section \ref{sec:random} that, at least for a large class of random matrices, the corresponding $\ell_1$-CMSVs are bounded away from zero with high probability if $m \geq c_1 k\log\frac{n}{k}$.

\section{$\ell_1$-constrained Minimal Singular Values of Random Matrices}\label{sec:random}

This section is devoted to analyzing the property of the $\ell_1$-CMSVs for the subgaussian ensemble.
We employ a recent estimate on the behavior of empirical processes involving subgaussian random variables \cite{mendelson2007subgaussian}.

Before we turn to the general empirical process result of \cite{mendelson2007subgaussian} developed by the delicate use of the powerful generic chaining idea, we need some notations and definitions. For a scalar random variable $X$, the Orlicz $\psi_2$ norm is defined as
\begin{eqnarray}
  \|X\|_{\psi_2} = \inf \left\{t > 0: \E \exp\left(\frac{|X|^2}{t^2}\right) \leq 2\right\}.
\end{eqnarray}
Markov's inequality immediately gives that $X$ with finite $\|X\|_{\psi_2}$ has subgaussian tail
\begin{eqnarray}
\Pr(|X| \geq t ) \leq 2 \exp(-ct^2/\|X\|_{\psi_2}).
\end{eqnarray}
The converse is also true, \emph{i.e.,} if $X$ has subgaussian tail $\exp(-t^2/K^2)$, then $\|X\|_{\psi_2} \leq c K$.

A random vector $\X \in \R^n$ is called \emph{isotropic and subgaussian} if $\E|\left<\X, \u\right>|^2 = \|\u\|_2^2$ and $\|\left<\X, \u\right>\|_{\psi_2} \leq L \|\u\|_2$ hold for any $\u \in \R^n$. A random vector $\X$ with independent subgaussian entries $X_1, \ldots, X_n$ is a subgaussian vector because \cite{mendelson2008embedding}
\begin{eqnarray}
\left\|\left<\X, \u\right>\right\|_{\psi_2} &\leq& c \sqrt{\sum_{i=1}^n \u_i^2 \|X_i\|_{\psi_2}^2} \nn\\
&\leq& c \max_{1\leq i \leq n}\|X_i\|_{\psi_2} \|\u\|.
\end{eqnarray}
Clearly, if in addition $\{X_i\}$ are centered and has unit variance, then $\X$ is also isotropic. In particular, the standard Gaussian vector on $\R^n$ and the sign vector with \emph{i.i.d.} $1/2$ Bernoulli entries are isotropic and subgaussian. Isotropic and subgaussian random vectors also include the vectors with the normalized volume measure on various convex symmetric bodies , for example, the unit balls of $\ell_p^n$ for $2 \leq p \leq \infty$ \cite{mendelson2007subgaussian}.\\



We reformulate the $\ell_1-$CMSV for sensing matrices with subgaussian entries using empirical processes. Suppose the entries of $A$ are \emph{i.i.d.} with subgaussian tails such that $\E \|A\u\|_2^2 = m\|\u\|_2^2$ for any $\u \in \R^n$. The rows of $A$ are denoted by $\{\a_i^T, i = 1,\ldots, m\}$. Denote $\H_{s}^{n} = \{\u \in \R^n: \|\u\|_2^2 = 1, \|\u\|_1^2 \leq s\}$. We note that $\rho_s(A/\sqrt{m}) > 1 - \epsilon$ is a consequence of
\begin{eqnarray}\label{eqn:supform}
&&\sup_{\u \in \H_s^n}\left|\frac{1}{m}\u^TA^TA\u - 1\right|\nn\\
&=&  \sup_{\u \in \H_s^n}\left|\frac{1}{m}\sum_{i=1}^m \left<\a_i, \u\right>^2 - 1\right| \leq \epsilon.
\end{eqnarray}
Define a class of functions parameterized by $\u$ as $\sF_s \df \{f_{\u}(\cdot) = \left<\u, \cdot\right>: \u \in \H_s^n\}$ and denote $P_m$ the empirical measure that puts equal mass at each of the $m$ random variables (observations) $\a_1, \ldots, \a_m$, \emph{i.e.},
\begin{eqnarray}
P_m(\cdot) = \frac{1}{m} \sum_{i=1}^m \delta_{\a_i}(\cdot)
\end{eqnarray}
with $\delta_{\x}(\cdot)$ the dirac measure that puts unit mass at $\x$. We realize that $\{\frac{1}{m}\sum_{i=1}^m \left<\a_i, \u\right>^2\}$ is the empirical process $\{P_m(f^2)\}_{f \in \sF_s}$. We slightly abuse notation and use $\E f^2$ to denote $\E f^2(\a)$. Then, our goal is to estimate
\begin{eqnarray}
  \E \sup_{f\in \sF_s}\left|P_m(f^2) - \E f^2\right|
\end{eqnarray}
and
\begin{eqnarray}\label{eqn:emprob}
  \Pr\left\{\sup_{f\in \sF_s}\left|P_m(f^2) - \E f^2\right|\right\},
\end{eqnarray}
a central topic of the study of empirical processes.

A key concept in studying general Gaussian processes as well as the empirical process $\{P_m(f^2)\}_{f \in \sF_s}$ is the $\gamma_p$ function we are going to define. We need some setup first. For any set $\sx$, an \emph{admissible sequence} is a sequence of increasing partitions $\{\Q_k\}_{k\geq 0}$ of $\sx$ such that $\mathrm{card}(\Q_0) = 1$ and $\mathrm{card}(\Q_k) = 2^{2^k}$ for $k \geq 1$. By a sequence of increasing partitions, we mean that every set in $\Q_{k+1}$ is contained in some set of $\Q_{k}$. We will use $Q_k(\x)$ to denote the unique set in partition $\Q_k$ that contains $\x$. The diameter of $Q_k(\x)$ is denoted by $\Delta(Q_k(\x))$. Then we have the following definition for $\gamma_p$ function associated with a metric space:
\begin{defi}\hskip -0.1cm \emph{\cite{talagrand2005generic}}
Suppose $(\sx, d)$ is a metric space and $p > 0$. We define
\begin{eqnarray}
  \gamma_p(\sx, d) &=& \inf \sup_{\x \in \sx} \sum_{k\geq 0} 2^{k/p} \Delta(Q_k(\x)),
\end{eqnarray}
where the infimum is taken over all admissible sequences.
\end{defi}

The importance of the $\gamma_p$ lies in its relationship with the behavior of Gaussian process indexed by a metric space when the metric is induced by the Gaussian process. More precise, suppose $\{X_{\x}\}_{\x \in \sx}$ is a Gaussian process indexed by the metric space $(\sx, d)$ with
\begin{eqnarray}\label{eqn:defofd}
  d(\x,\y) = (\E(X_{\x} - X_{\y})^2)^{1/2},
\end{eqnarray}
then we have
\begin{eqnarray}\label{eqn:gaussgamma}
  c \gamma_2(\sx, d) \leq \E \sup_{\x \in \sx} X_{\x} \leq C \gamma_2(\sx, d)
\end{eqnarray}
for some numerical constants $c$ and $C$. The upper bound (the generic chaining bound) was first established by Fernique \cite{fernique1975regularite} and the lower bound is obtained by Talagrand using majorizing measures \cite{talagrand1987regularity}. The rather difficult concept of majorizing measures has been considerably simplified through the notion of ``generic chaining", an idea that dates back to Kolmogorov and is greatly advanced in recently years by Talagrand \cite{talagrand2005generic}.

With these preparations, we present the major result of \cite{mendelson2007subgaussian}:
\begin{thm}\hskip -0.1cm\emph{\cite{mendelson2007subgaussian}}\label{thm:empirical}
Let $\{\a, \a_i, i = 1,\ldots,m\} \subset \R^n$ be \emph{i.i.d.} random vectors which induce a measure $\mu$ on $\R^n$, and $\sF$ be a subset of the unit sphere of $L_2(\R^n, \mu)$ with $\mathrm{diam}(\sF, \|\cdot\|_{\psi_2}) = \alpha$. Then there exist absolute constants $c_1, c_2, c_3$ such that for any $\epsilon > 0$ and $m \geq 1$ satisfying
\begin{eqnarray}
  m \geq c_1 \frac{\alpha^2 \gamma_2^2(\sF, \|\cdot\|_{\psi_2})}{\epsilon^2},
\end{eqnarray}
with probability at least $1 - \exp(-c_2\epsilon^2 m/\alpha^4)$,
\begin{eqnarray}
  \sup_{f\in \sF}\left|\frac{1}{m} \sum_{k=1}^m f^2(\a_k) - \E f^2(\a)\right| \leq \epsilon.
\end{eqnarray}
Furthermore, if $\sF$ is symmetric, we have
\begin{eqnarray}
\hskip -1cm &&\E \sup_{f \in \sF} \left|\frac{1}{m} \sum_{k=1}^m f^2(\a_k) - \E f^2(\a)\right| \nn\\
\hskip -1cm &&\leq c_3 \max\left\{\alpha \frac{\gamma_2(\sF, \|\cdot\|_{\psi_2})}{\sqrt{m}}, \frac{\gamma_2^2(\sF, \|\cdot\|_{\psi_2})}{m}\right\}.
\end{eqnarray}
\end{thm}

We apply Theorem \ref{thm:empirical} to estimate the $\ell_1$-CMSV. Consider the function set $\sF = \sF_s = \{f_{\u}(\cdot) = \left<\u, \cdot\right>: \|\u\|_2^2 = 1, \|\u\|_1^2 \leq s\}$. Assume $\a \in \R^n$ is isotropic and subgaussian. As a consequence of the isotropy of $\a$ and $\|\u\|_2 = 1$, we get $\sF_s$ is a subset of the unit sphere of $L_2(\R^n, \mu)$. The symmetry of $\sF_s$ yields
\begin{eqnarray}
\alpha &=& \mathrm{diam}(\sF_s, \|\cdot\|_{\psi_2})\nn\\
& =& 2 \sup_{\u \in \H_s^n} \E \left<\u,\a\right>^2 = 2.
\end{eqnarray}

Now the key is to compute $\gamma_2(\sF_s, \|\cdot\|_{\psi_2})$. Due to \eqref{eqn:gaussgamma}, the problem reduces to the computation of $\E \sup_{\u \in \H_s^n} X_u$ (actually an upper bound suffices), where $\{X_{\u}\}_{\u \in \H_s^n}$ is the canonical Gaussian process:
\begin{eqnarray}
X_{\u} = \left<\g, \u\right>, \ \g \sim \N(0, \I_n), \u \in \H_s^n.
\end{eqnarray}
Clearly, we have
\begin{eqnarray}
\gamma_2(\sF_s, \|\cdot\|_{\psi_2}) &\leq& c\ \E \sup_{\u \in \H_s^n} \left<\g, \u\right> \nn\\
&\leq& c\ \E \|\u\|_1\|\g\|_\infty \nn\\
&\leq& c\ \sqrt{s\log n}.
\end{eqnarray}

As a consequence, we have the following theorem:
\begin{thm}
Let the rows of the sensing matrix $A$ be \emph{i.i.d.} subgaussian and isotropic random vectors. Then there exists constants $c_1, c_2, c_3$ such that for any $\epsilon > 0$ and $m \geq 1$ satisfying
\begin{eqnarray}
  m \geq c_1 \frac{s \log n}{\epsilon^2},
\end{eqnarray}
we have
\begin{eqnarray}
  \E |1 - \rho_s(A)| \leq c_2\epsilon,
\end{eqnarray}
and
\begin{eqnarray}
  \Pr\{1 - \epsilon \leq \rho_s(A) \leq 1 + \epsilon\} \geq  1 - \exp(-c_3 \epsilon^2m).
\end{eqnarray}
\end{thm}

This theorem says that at least for subgaussian ensembles (including the Gaussian ensemble and the Bernoulli ensemble), the $\ell_1$-CMSV bounds are as tight as the RIC bounds.

\section{Computation of the $\ell_1$-CMSVs}\label{sec:computation}
 In this section, we first describe two algorithms to compute a lower bound on the maximal $k$ such that the sufficient condition \eqref{nullspaceproperty} is satisfied. This gives a way to verify that the $\ell_1$ recovery is exact in the noiseless setting. The second part of this section is devoted to the computation of $\ell_1$-CMSV and its lower bound.
 \subsection{Verifying the Sufficient Condition for Exact $\ell_1$ Recovery}
 Using the $\ell_1$-sparsity level to verify the sufficient condition \eqref{nullspaceproperty} (refer to Section \ref{sec:bound}) is formulated as the following optimization problem:
 \begin{eqnarray}
   \min_{\z: A\z = 0} \frac{1}{4}\frac{\|\z\|_1^2}{\|\z\|_2^2},
 \end{eqnarray}
 or equivalently,
  \begin{eqnarray}
   \max_{\z} 4\|\z\|_2^2 \text{\  s.t.\ } A\z = 0, \|\z\|_1 \leq 1.
 \end{eqnarray}
 Unfortunately, this later optimization problem, which maximizes the $\ell_2$ norm over a polyhedron, is NP-hard \cite{Bodlaender1990Normmaximization}. By defining $Z = \z\z^T$ and dropping the rank constraint, we instead use the following semidefinite relaxation  to produce a lower bound:
 \begin{eqnarray}\label{eqn:l2}
   \hskip -0.5cm \mathrm{(L_2):}&&\max_{Z: Z \succeq 0} 4 \tr(Z) \nn\\
   \hskip -0.5cm &&\text{\  s.t.\ } \tr(AZA^T) = 0, \|Z\|_1 \leq 1,
 \end{eqnarray}
 where $\|Z\|_1$ is the sum of absolute values of all elements in $Z$.

 Another relaxation based on the $\ell_\infty$ norm is to solve the following optimization problem (refer to \eqref{eqn:maxfrac1inf}):
 \begin{eqnarray}\label{eqn:linf}
\mathrm{(L_\infty):}\ \ \ \max_{\z} 2\|\z\|_\infty \text{\ s.t.\ } A\z = 0, \|\z\|_1 \leq 1,
 \end{eqnarray}
which is solved by the following $n$ linear programs:
 \begin{eqnarray}
\max_{\z} 2\z_i \text{\ s.t.\ } A\z = 0, \|\z\|_1 \leq 1, i = 1,\ldots, n.
 \end{eqnarray}
We observe this linear program subproblem is actually the dual of the  linear program subproblem used in \cite{Juditsky2010Verifiable} to compute $\alpha_1(A,\beta)$ when $\beta = \infty$. In this paper, the linear program subproblems are implemented using the primal-dual algorithm detailed in Chapter 11 of \cite{boyd2004convex}. This algorithm produces results comparable to those in \cite{Juditsky2010Verifiable} but is significantly faster.

 \subsection{Computing $\ell_1$-CMSV}
An advantage of using the $\ell_1$-CMSV as a measure of the ``goodness" of a sensing matrix is the \emph{relative} ease of its computation. The computation of $\ell_1$-CMSV is equivalent to
\begin{eqnarray}\label{opt_l1cmsv}
\min_{\x \in \R^n} \|A\x\|_2 \text{\ \ s.t.\ }\|\x\|_1 \leq \sqrt{s},\ \|\x\|_2 = 1.
\end{eqnarray}
Unfortunately, the above optimization is not convex because of the $\ell_2$ constraint $\|\x\|_2 = 1$. However, many tools at our disposal can deal with the continuous problem \eqref{opt_l1cmsv}, for example, the Lagrange multiplier or  the Karush-Kuhn-Tucker condition \cite{Nash1996Programming}. We will present an interior point algorithm to directly compute an approximate numerical solution of \eqref{opt_l1cmsv}. Since the optimization problem \eqref{opt_l1cmsv} is not convex, there is no guarantee that the solution of the algorithm are the true minima. Thus, we will also present a convex program to compute a lower bound on $\ell_1$-CMSV.\\

The interior point (IP) method provides a general approach to efficiently solve the following general constrained optimization problem:
\begin{eqnarray}\label{gp}
\min_{\z \in \R^n} F(\z) \text{\ s.t. \ } f(\z) \leq 0,  \ g(\z) = 0.
\end{eqnarray}
The basic idea is to construct and solve a sequence of penalized optimization problems with equality constraints:
\begin{eqnarray}\label{penalized_op}
  &&\ \ \ \ \ \min_{\z, \sigma} F(\z) - \mu \sum_i \log(\sigma_i) \nonumber\\
  && \text{\ s.t.\ } f(\z) + \sigma = 0, \ g(\z) = 0.
\end{eqnarray}
By using either a Newton step, which tries to solve the Karush-Kuhn-Tucker equations \cite{Nash1996Programming}, or a conjugate gradient step using trust regions to solve the penalized problem \eqref{penalized_op} in each iteration, the interior point approach efficiently generates a sequence of solutions that converge to the solution of \eqref{gp}. Refer to \cite{byrd2000a, byrd1999an, waltz2006an} for more information on this approach.

However, the interior point approach assumes that the objective and constraint functions have continuous second order derivatives, which is not satisfied by the constraint $\|\z\|_1 - \sqrt{s} \leq 0$. We address the non-differentiability of $f(\z) = \|\z\|_1 - \sqrt{s}$ by defining $\z = \z^+ - \z^-$ with $\z^+ = \max(\z, \zero) \geq 0$ and $\z^- = \max(-\z, \zero) \geq 0$, which leads to the following augmented optimization:
\begin{eqnarray}\label{ip2}
  &&\text{IP:\ \ }\ \min_{\z^+, \z^- \in \R^n} (\z^+ - \z^-)^TA^TA(\z^+ - \z^-)\nonumber\\
  && \text{subject\ } \text{to\ } \sum_i \z^+_i + \sum_i \z^-_i - s \leq 0, \nonumber\\
  && \ \ \ \ \ \ \ \ \ \ \ \ (\z^+ - \z^-)^T(\z^+ - \z^-) = 1, \nonumber\\
  && \ \ \ \ \ \ \ \ \ \ \ \ \z^+ \geq 0 ,\ \z^- \geq 0.
\end{eqnarray}
This algorithm is employed in \cite{Sen2010ofdm} to design the transmitting waveform of an OFDM radar for optimal detection and estimation performance.
\\

We briefly describe a semidefinite relaxation (SDR) approach to compute a lower bound on $\ell_1$-CMSV. A similar method was employed in \cite{dAspremont2007sparsePCA} to compute an upper bound on sparse variance maximization using the \emph{lifting procedure} for semidefinite programming \cite{schrijver1991cones, alizadeh1995interior, oustry1999semidefinite}. Defining $Z = \z\z^T$ and dropping the rank constraint transform problem \eqref{opt_l1cmsv} into the following form:
\begin{eqnarray}\label{sr}
 \text{SDR:\ \ } \min_{Z \succeq 0} &&\tr(A^TAZ)\nonumber\\
  \text{s.t.\ }&&\|Z\|_1 \leq s, \tr(Z) = 1.
\end{eqnarray}
Now SDR is a semidefinite programming problem. For a small size problem, a global minimum can be achieved at high precision using SEDUMI \cite{Sturm1999Sedumi}, SDPT3 \cite{Tutuncu2003SDPT3} or CVX \cite{Grant2009CVX}. However, for relatively large $n$, the interior point algorithm makes the memory requirement prohibitive (see \cite{dAspremont2007sparsePCA} for more discussion). In this paper, we do not consider more efficient implementations of the SDR.

\section{Numerical Simulations}\label{sec:simulations}
We use numerical simulations to assess the effectiveness and efficiency of the algorithms presented in Section \ref{sec:computation}. Except for the JN algorithm in Table \ref{tbl:Hadamard1024}, all other experiments were performed on a platform with a Pentium D CPU@3.40GHz, 2GB RAM, and a Windows XP operating system.

\subsection{Verification of Sufficient Conditions}We first examine the $\mathrm{L_2}$ \eqref{eqn:l2} and $\mathrm{L_\infty}$ \eqref{eqn:linf} algorithms for verifying the sufficient condition \eqref{nullspaceproperty}. These two algorithms are compared with the two algorithms d'AE and JN proposed in \cite{dAspermont2010Nullspace} and \cite{Juditsky2010Verifiable}, respectively. We name the two algorithms d'AE and JN using the abbreviations of the authors' names. Recall that $k^*$ is defined as the maximal $k$ such that \eqref{nullspaceproperty} is satisfied. In Table \ref{tbl:Bernoulli40}, we show the lower bounds on $k^*$ for a small size Bernoulli matrix with $n = 40$ computed by $\mathrm{L_2}$, $\mathrm{L_\infty}$, d'AE and JN. The algorithms of d'AE and JN are provided by the authors for free download online.
%

\begin{table}[h!t]
\caption{Comparison of different verification algorithms for a Bernoulli matrix with leading dimension $n = 40$.}
\begin{center}
\begin{tabular}{||l||l|l|l|l||l|l|l|l||}
\hline
\multirow{2}{*}{$m$} & \multicolumn{4}{|c||}{lower bounds on $k^*$} & \multicolumn{4}{|c||}{CPU time (s)}\\
\cline{2-5}\cline{6-9}
& $\mathrm{L_1}$ & $\mathrm{L_\infty}$ & d'AE & JN &  $\mathrm{L_1}$ & $\mathrm{L_\infty}$ & d'AE & JN \\
\hline
\hline
20 & 1 & 1 & 2 & 1 & 14.29 & 0.41 & 1040.40 & 4.59\\
\hline
24 & 1 & 2 & 2 & 2 & 16.11 & 0.38 & 694.20 & 0.69\\
\hline
28 & 2 & 3 & 3 & 3 & 15.12 & 0.37 & 710.90 & 16.20\\
\hline
32 & 2 & 3 & 4 & 3 & 15.43 & 0.37 & 894.08 & 2.45\\
\hline
\end{tabular}\label{tbl:Bernoulli40}
\end{center}
\end{table}

\begin{table}[h!t]
\caption{Comparison of $\mathrm{L_\infty}$ and JN for a Hadamard matrix with leading dimension $n = 256$.}
\begin{center}
\begin{tabular}{||l||l|l||l|l||}
\hline
\multirow{2}{*}{$m$} & \multicolumn{2}{|c||}{lower bounds on $k^*$} & \multicolumn{2}{|c||}{CPU time (s)}\\
\cline{2-3}\cline{4-5}
& $\mathrm{L_\infty}$ & JN & $\mathrm{L_\infty}$ & JN\\
\hline
\hline
25 & 1 & 1 & 3 & 35\\
\hline
51 & 2 & 2 & 6 & 70\\
\hline
76 & 3 & 3 & 7 & 102\\
\hline
102 & 4 & 4 & 9 & 303\\
\hline
128 & 5 & 5 & 9 & 544\\
\hline
153 & 7 & 7 & 13 & 310\\
\hline
179 & 9 & 9 & 15 & 528\\
\hline
204 & 12 & 12 & 18 & 1333\\
\hline
230 & 19 & 18 & 18 & 435\\
\hline
\end{tabular}\label{tbl:Hadamard256}
\end{center}
\end{table}

\begin{table}[h!t]
\caption{Comparison of $\mathrm{L_\infty}$ and JN for a Gaussian matrix with leading dimension $n = 256$.}
\begin{center}
\begin{tabular}{||l||l|l||l|l||}
\hline
\multirow{2}{*}{$m$} & \multicolumn{2}{|c||}{lower bnd on $k^*$} & \multicolumn{2}{|c||}{CPU time (s)}\\
\cline{2-3}\cline{4-5}
& $\mathrm{L_\infty}$ & JN & $\mathrm{L_\infty}$ & JN\\
\hline
\hline
25 & 1 & 1 & 6 & 91\\
\hline
51 & 2 & 2 & 8 & 191\\
\hline
76 & 3 & 3 & 10 & 856\\
\hline
102 & 4 & 4 & 13 & 5630\\
\hline
128 & 4 & 5 & 16 & 5711\\
\hline
153 & 6 & 6 & 20 & 1381\\
\hline
179 & 7 & 7 & 24 & 3356\\
\hline
204 & 10 & 10 & 25 & 10039\\
\hline
230 & 13 & 14 & 28 & 8332\\
\hline
\end{tabular}\label{tbl:Gaussian256}
\end{center}
\end{table}

\begin{table}[h!t]
\caption{Comparison of $\mathrm{L_\infty}$ and JN for Gaussian and Hadamard matrices with leading dimension $n = 1024$. In the column head, ``G" represents Gaussian matrix and ``H" represents Hadamard matrix.}
\begin{center}
\begin{tabular}{||l||l|l|l||l|l|l||}
\hline
\multirow{3}{*}{$m$} & \multicolumn{3}{|c||}{lower bounds on $k^*$} & \multicolumn{3}{|c||}{CPU time (s)}\\
\cline{2-7}
& $\mathrm{L_\infty}$(H) & $\mathrm{L_\infty}$(G)& JN(G) & $\mathrm{L_\infty}$(H)  & $\mathrm{L_\infty}$(G) & JN(G)\\
\hline
\hline
102 & 3 & 2 & 2 & 182& 136 & 457\\
\hline
204 & 4 & 4 & 4 & 501 & 281 & 1179\\
\hline
307 & 6 & 6 & 6 & 872 & 510 & 2235\\
\hline
409 & 8 & 7 & 7 & 1413 & 793 & 3659\\
\hline
512 & 11 & 10 & 10 & 1914 & 990 & 5348\\
\hline
614 & 14 & 12 & 12 & 1362 & 1309 & 7156\\
\hline
716 & 18 & 15 & 15 & 1687 & 1679 & 9446\\
\hline
819 & 24 & 20 & 21 & 1972 & 2033 & 12435\\
\hline
921 & 37 & 29 & 32 & 2307 & 2312 & 13564\\
\hline
\end{tabular}\label{tbl:Hadamard1024}
\end{center}
\end{table}
In the next set of experiments, we compare lower bounds on $k^*$ computed by $\mathrm{L_\infty}$ and JN, respectively, for $n = 256$. In this case, both the semidefinite relaxation in this paper and that in \cite{dAspermont2010Nullspace} are too time and memory consuming to compute. The lower bounds and execution times are shown in Table \ref{tbl:Hadamard256} and \ref{tbl:Gaussian256} for a Hadamard and a Gaussian matrix, respectively.

Table \ref{tbl:Hadamard1024} shows the results of $\mathrm{L_\infty}$ and JN for a Hadamard matrix with leading dimension $n = 1024$. Note the lower bounds computed by JN and the CPU times of JN in table \ref{tbl:Hadamard1024} are extracted from \cite{Juditsky2010Verifiable}. We were not able to carry out the computation of JN within reasonable time in our platform.\\

From Table \ref{tbl:Bernoulli40}, we see that for $n = 40$, d'AE performs the best, and $\mathrm{L_\infty}$ and JN give exactly the same results. However, d'AE is very slow in general. For example, even with a first order implementation, the d'AE takes more than 37 hours for matrices of size $350\times 500$, while the $\mathrm{L_\infty}$ takes less than 3 minutes to finish the computation. From Table \ref{tbl:Hadamard256}, \ref{tbl:Gaussian256} and \ref{tbl:Hadamard1024}, we see that $\mathrm{L_\infty}$ and JN produces comparable results. However, our $\mathrm{L_\infty}$ algorithm performs much faster than the JN algorithm. Because the two algorithms solve $n$ linear program subproblems that are dual to each other, they should yield exactly the same results. However, we observe that sometimes the upper bound and lower bound on \emph{the lower bound of $k^*$} computed by JN does not coincide. The difference in speed might come from the implementation. Our implementation of the linear sub-program employs the primal-dual approach detailed in \cite{boyd2004convex} while \cite{Juditsky2010Verifiable} uses the commercial LP solver \verb"mosekopt" \cite{Andersenmosek}.

\subsection{Computation of $\ell_1$-CMSV} We next report the test results of the IP and SDR algorithms for computing the $\ell_1$-CMSV and its lower bound, respectively. The interior point algorithms IP is implemented using the MATLAB\textsuperscript{\textregistered} function \emph{fmincon}. The SDR is solved using CVX \cite{Grant2009CVX} with the default SDPT3 solver.

We fist test IP and SDR on a Gaussian matrix $A \in \R^{20\times 60}$ for $s = 5$. Due to the existence of local minima, we need to run IP several times and select the minimal function value among all the trials as the $\ell_1$-CMSV. Fifty random initial points on the unit sphere in $\R^{60}$ are generated for IP.  The SDR only runs once. The results are shown in Table \ref{tbl:A60}. In this example, SDR over-relaxes the problem and produces a zero $\ell_1$-CMSV.


\begin{table}[h!]
\centering
\caption{Comparison of IP and SDR for a Gaussian matrix.}
\label{tbl:A60}
\begin{tabular}{|l|l|l|l|l|}
  \hline
  & $\min F(\z^*)$ & $\mean F(\z^*)$ & $\std F(\z^*)$ & mean time (s)  \\
  \hline
  \hline
  IP & 0.0666 & 0.7133 & 0.3661 & 2.8903  \\
  \hline
  SDR & 0.0000 & 0.0000 & N/A & 53.1583 \\
  \hline
\end{tabular}
\end{table}
\begin{table}[h!t]
\centering
\caption{IP for a Bernoulli matrix $A \in \R^{50\times 500}.$}
\label{tbl:A500}
\begin{tabular}{|l|l|l|l|l|}
  \hline
  & $\min F(\z^*)$ & $\mean F(\z^*)$ & $\std F(\z^*)$ & mean time (s) \\
  \hline\hline
  IP & 0.000371 & 0.007472  & 0.004545 & 123.6456 \\
  \hline
\end{tabular}
\end{table}
The IP is also tested for a Bernoulli matrix $A \in \R^{50\times 500}$ with results shown in Table \ref{tbl:A500}. The CVX implementation of SDR takes too much memory to run for $n = 500$.\\

\begin{figure}
\centering
\includegraphics[trim = 18mm 0mm 30mm 0mm, clip, width = 0.48\textwidth]{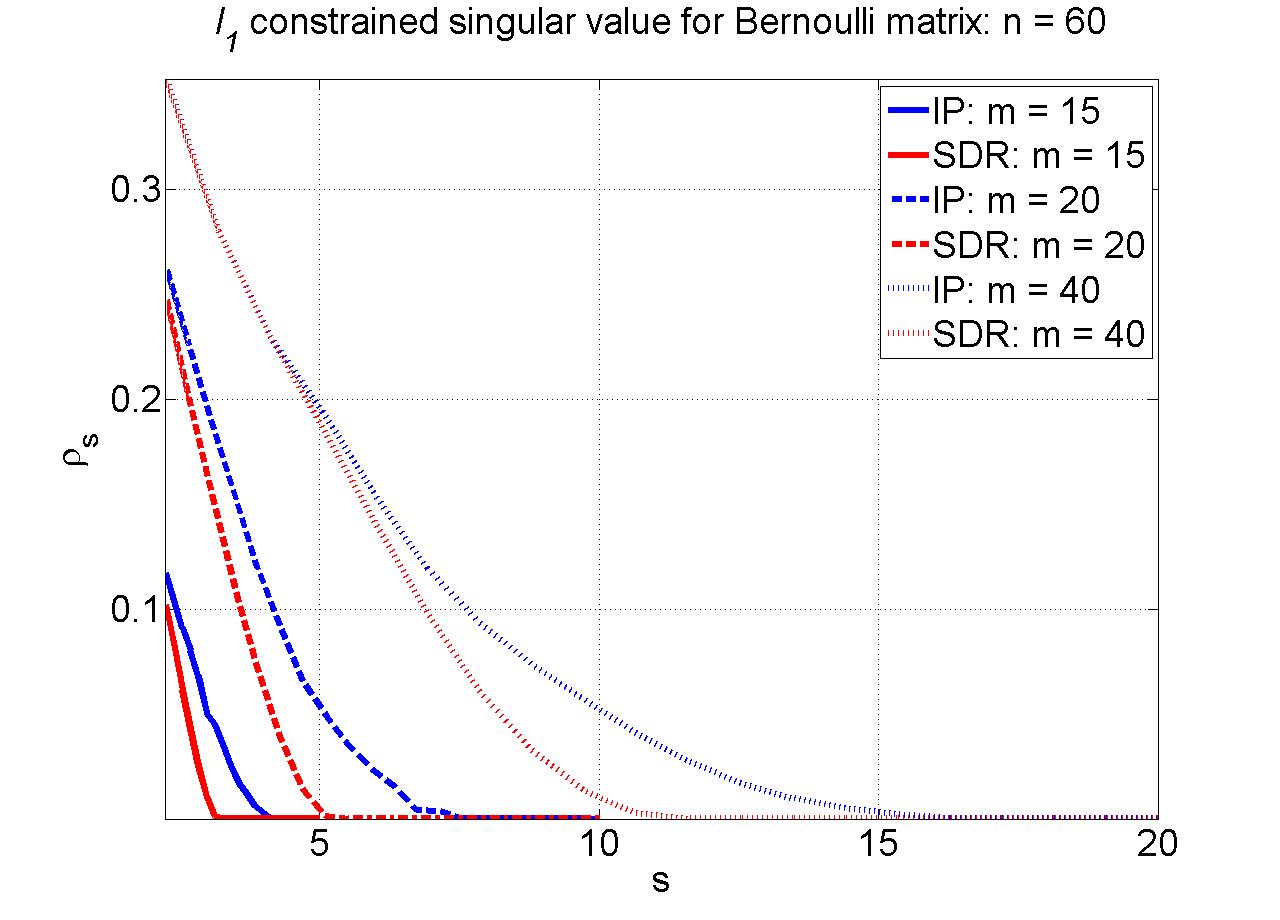}
\caption{$\ell_1$-CMSV $\rho_s$ and its bound as a function of $s$ for Bernoulli matrix with $n = 60$ and $m = 10, 20, 40$.}
\label{fig:svary60}
\end{figure}

\begin{figure}
\centering
\includegraphics[trim = 15mm 0mm 25mm 0mm, clip, width = 0.5\textwidth]{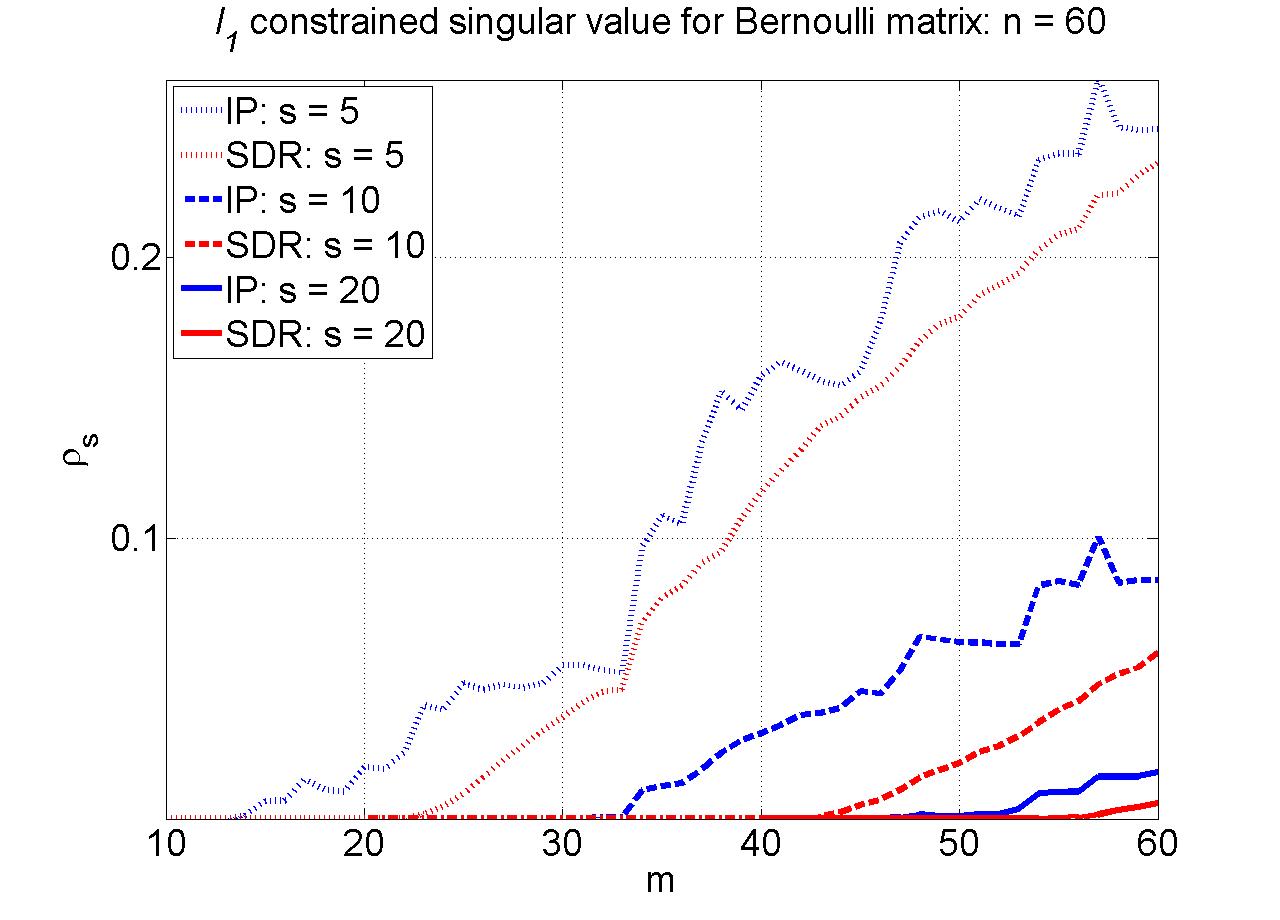}
\caption{$\ell_1$-CMSV $\rho_s$ and its bound as a function of $m$ for Bernoulli matrix with $n = 60$ and $s = 5, 10, 20$.}
\label{fig:mvary60}
\end{figure}

We compare the $\ell_1$-CMSVs $\rho_s$ and their bounds as a function of $s$ computed by IP and SDR, respectively, for Bernoulli random matrices. We consider a small-scale problem with $n = 60$ and $m = 10, 20 ,40$. A matrix $B \in \R^{40\times 60}$ with entries $\{+1, -1\}$ following $\frac{1}{2}$ Bernoulli distribution is generated. For $m = 10, 20, 40$, the corresponding Bernoulli matrix $A$ is obtained by taking the first $m$ rows of $B$. The columns of $A$ are then normalized to have unit norm. The normalization implies that $\rho_s \leq \rho_1  = 1$. The IP uses $30$ random initial points. As illustrated in Figure \ref{fig:svary60}, the $\ell_1$-CMSVs and their bounds decrease very fast as $s$ increases. For fixed $s$, increasing $m$ generally (but not necessarily, as shown in Figure \ref{fig:mvary60}) increases the $\ell_1$-CMSV and their bounds.

In Figure \ref{fig:mvary60}, the $\ell_1$-CMSV $\rho_s$ is plotted as a function of $m$ with varying parameter values: $s = 5, 10$ and $20$. With $s$ fixed, the two algorithms (IP and SDR) are run for $A \in \R^{m\times n}$, with $m$ increasing from $2s$ to $n = 60$. For each $m$, the construction of $A$ follows the procedure described in the previous paragraph. The discrete nature of adding rows to $A$ while increasing $m$ makes the curves in Figure \ref{fig:mvary60} not as smooth as those in Figure \ref{fig:svary60}. The $\rho_s$ increases with $m$ in general, but local decreases do happen. The gap between values computed by IP and SDR is also clearly seen for medium $s$.

\section{Conclusions}\label{sec:conclusions}

In this paper, a new measure of a sensing matrix's incoherence, the $\ell_1$-CMSV, is proposed to quantify the stability of sparse signal reconstruction. It is demonstrated that the reconstruction errors of the Basis Pursuit, the Dantzig selector, and the LASSO estimator are concisely bounded using the $\ell_1$-CMSV. A generic chaining argument shows that the $\ell_1$-CMSV is bounded away from zero with high probability for the subgaussian ensemble, as long as the number of measurements is relatively large. One interior point program and one semidefinite program are presented to compute the $\ell_1$-CMSV and its lower bound, respectively. Numerical simulations assess the algorithms' performance. The $\ell_1$-CMSV provides a computationally amenable measure of incoherence that can be used for optimal design.

As a by product, two algorithms are designed to verify the sufficient conditions guaranteeing the uniqueness of $\ell_1$-based recovery. The $\ell_\infty$ relaxation based algorithm is shown to produce comparable results with the state-of-the-art algorithms, and performs much faster.
\appendices
\section{Proof of Theorem \ref{thm:errorbd}}\label{sec:pf:errorbd}
In this appendix, we derive the error bounds presented in Theorem \ref{thm:errorbd}.

\begin{proof}[Proof of Theorem \ref{thm:errorbd}] We strictly follow the two-step procedure expounded in Section \ref{sec:bound}.\\

1) In order to establish the $\ell_1$-sparsity of the error vector in the first step, we suppose $S = \supp(\x)$ and $|S| = \|\x\|_0 = k$. Define the error vector $\h = \hx - \x$. For any vector $\z \in \R^n$ and any index set $S \subseteq \{1,\ldots,n\}$, we use $\z_S \in \R^{|S|}$ to represent the vector whose elements are those of $\z$ indicated by $S$.\\

We first deal with the BP and the DS. As observed by Cand\'es in \cite{Candes2008RIP},  the fact that $\|\hx\|_1 = \|\x + \h\|_1$ is the minimum among all $\z$ satisfying the constraints in \eqref{bp} and \eqref{ds}, \emph{together with} the fact that the true signal $\x$ satisfies the constraints as required by the conditions imposed on the noise in Theorem \ref{thm:errorbd}, imply that $\|\h_{S^c}\|_1$ cannot be very large. To see this, we observe that
\begin{eqnarray}\label{x_min}
  \|\x\|_1 &\geq& \|\x + \h\|_1\nonumber\\
  & = & \sum_{i\in S}|\x_i + \h_i| + \sum_{i\in S^c}|\x_i + \h_i| \nonumber\\
  &\geq& \|\x_S\|_1 - \|\h_S\|_1 + \|\h_{S^c}\|_1\nonumber\\
  & = & \|\x\|_1 - \|\h_S\|_1 + \|\h_{S^c}\|_1.
\end{eqnarray}
Therefore, we obtain $\|\h_{S^c}\|_1 \leq \|\h_S\|_1$, which leads to
\begin{eqnarray}\label{h1h2}
\|\h\|_1  &=& \|\h_S\|_1 + \|\h_{S^c}\|_1\nonumber\\
 &=&  2\|\h_S\|_1 \nonumber\\
 &\leq& 2\sqrt{k}\|\h_S\|_2 \nonumber \\
 &\leq& 2\sqrt{k}\|\h\|_2,
\end{eqnarray}
where for the next to the last inequality we used the Cauchy-Schwart inequality. Inequality \eqref{h1h2} is equivalent to
\begin{eqnarray}\label{s_1}
s(\h) \leq 4k.
\end{eqnarray}
\\

We now continue to establish the $\ell_1$ sparsity of the error vector for the LASSO \eqref{lasso}. We borrow ideas from \cite{candes2009lowrank} (see also \cite{bickel2009simultaneous}). Since the noise $\w$ satisfies $\|A^T\w\|_\infty \leq \kappa\lambda_n\sigma$ for some small $\kappa > 0$ and $\hx$ is a solution to \eqref{lasso}, we have
\begin{eqnarray*}
\frac{1}{2}\|A\hx - \y\|_2^2 + \lambda_n\sigma \|\hx\|_1 \leq \frac{1}{2} \|A\x - \y\|_2^2 + \lambda_n\sigma\|\x\|_1.
\end{eqnarray*}
Consequently, substituting $\y = A\x + \w$ yields
\begin{eqnarray*}
\lambda_n\sigma \|\hx\|_1 &\leq& \frac{1}{2} \|A\x - \y\|_2^2 - \frac{1}{2}\|A\hx - \y\|_2^2 + \lambda_n\sigma \|\x\|_1 \nn\\
&=& \frac{1}{2}\|\w\|_2^2 - \frac{1}{2}\|A(\hx-\x) - \w\|_2^2 + \lambda_n\sigma\|\x\|_1\nn\\
& = & \frac{1}{2}\|\w\|_2^2 - \frac{1}{2}\|A(\hx-\x)\|_2^2\nn\\
  && \ \ \ + \left<A(\hx-\x), \w\right> - \frac{1}{2}\|\w\|_2^2 + \lambda_n\sigma\|\x\|_1\nn\\
&\leq & \left<A(\hx-\x), \w\right> + \lambda_n\sigma\|\x\|_1\nn\\
& = & \left<\hx - \x, A^T\w\right> + \lambda_n\sigma\|\x\|_1.
\end{eqnarray*}
Using the Cauchy-Swcharz type inequality, we get
\begin{eqnarray*}
\lambda_n\sigma \|\hx\|_1 &\leq & \|\hx-\x\|_1\|A^T\w\|_\infty + \lambda_n\sigma\|\x\|_1\nn\\
& = & \kappa \lambda_n\sigma \|\h\|_1 + \lambda_n\sigma\|\x\|_1,
\end{eqnarray*}
which leads to
\begin{eqnarray*}
  \|\hx\|_1 &\leq & \kappa\|\h\|_1 + \|\x\|_1.
\end{eqnarray*}
Therefore, similar to the argument in \eqref{x_min} we have
\begin{eqnarray*}
 &&\|\x\|_1 \nn\\
 &\geq& \|\hx\|_1 - \kappa\|\h\|_1\nn\\
 & = & \|\x + \h_{S^c} + \h_S\|_1- \kappa\left(\|\h_{S^c} + \h_S\|_1\right) \nonumber\\
  &\geq& \|\x + \h_{S^c} \|_1 - \|\h_S\|_1 - \kappa\left(\|\h_{S^c}\|_1 + \|\h_S\|_1\right) \nonumber\\
  & = & \|\x\|_1 + (1-\kappa)\|\h_{S^c}\|_1 - (1+\kappa)\|\h_S\|_1,
\end{eqnarray*}
where $S = \supp(\x)$.
Consequently, we have
\begin{eqnarray*}
  \|\h_{S^c}\|_1 &\leq& \frac{1+\kappa}{1-\kappa}\|\h_S\|_1.
\end{eqnarray*}
Therefore, an argument similar to the one leading to \eqref{h1h2} yields
\begin{eqnarray}\label{lassoh1h2}
\|\h\|_1 \leq \frac{2}{1-\kappa} \sqrt{k} \|\h\|_2,
\end{eqnarray}
or equivalently,
\begin{eqnarray}
  s(\h) &\leq& \frac{4k}{(1-\kappa)^2}.
\end{eqnarray}
\\

2) We now turn to obtain an upper bound on $\|A\h\|_2$. For the BP \eqref{bp}, this is trivial because both $\x$ and $\hx$ satisfy constraint $\|\y-A\z\| \leq \epsilon$ in \eqref{bp}. The triangle inequality yields
\begin{eqnarray}\label{Ah}
\|A\h\|_2 &=& \|A(\hx - \x)\|_2 \nonumber\\
&\leq& \|A\hx - \y\|_2 + \|\y - A\x\|_2 \nonumber\\
&\leq& 2 \epsilon.
\end{eqnarray}
It then follows from Definition \ref{def:l1cmsv} that
\begin{eqnarray}
  \rho_{4k} \|\h\|_2  &\leq & \|A\h\|_2 \leq 2\epsilon.
\end{eqnarray}
Hence, we get
\begin{eqnarray}
  \|\hx - \x\|_2 &\leq& \frac{2\epsilon}{\rho_{4k}}.
\end{eqnarray}
\\

For the DS \eqref{ds}, as shown in \cite{Candes2007Dantzig}, the condition on noise $\|A^T\w\|_\infty \leq \lambda_n \sigma$ and the constraint in the Dantzig selector \eqref{ds} yield
\begin{eqnarray}
  \|A^TA\h\|_\infty &\leq& 2\lambda_n \sigma
\end{eqnarray}
because
\begin{eqnarray}
  &&A_j^T(\w - \hat{\r}) = A_j^T[(\y-A\x)-(\y-A\hx)] \nonumber\\
  &=& A_j^T(A\hx - A\x) = A_j^TA\h,
\end{eqnarray}
where $\hat{\r} = \y - A\hx$ is the residual corresponding to the Dantzig selector solution $\hx$.
Therefore, we obtain an upper bound on $\|A\h\|_2^2$ as follows:
\begin{eqnarray}\label{ds_Ah_ubd}
  \h^TA^TA\h &=& |\sum_{i=1}^n\h_i(A^TA\h)_i|\nonumber\\
  & \leq &  \sum_{i=1}^n|\h_i|\cdot|(A^TA\h)_i|\nonumber\\
  & \leq & 2\lambda_n\sigma \|\h\|_1.
\end{eqnarray}
Equation \eqref{ds_Ah_ubd}, the definition of $\rho_{4k}$, and equation \eqref{h1h2} together yield
\begin{eqnarray}
  \rho_{4k}^2 \|\h\|_2^2 &\leq& \h^TA^TA\h \nonumber\\
  &\leq& 2\lambda_n\sigma \|\h\|_1 \nonumber\\
  &\leq& 4\lambda_n \sqrt{k}\sigma \|\h\|_2.
\end{eqnarray}
We conclude that
\begin{eqnarray}
  \|\hx - \x\|_2 &\leq& \frac{4\sqrt{k}}{\rho_{4k}^2} \lambda_n \sigma.
\end{eqnarray}
\\

Now we establish an upper bound on $\|A\h\|_2^2$ for the LASSO \eqref{lasso} using a  procedure similar to the one used for the DS given above. We need to establish a bound on
\begin{eqnarray}\label{lassoastara}
 && \|A^TA\h\|_\infty\nn\\
&\leq& \|A^T(\y-A\x)\|_\infty + \|A^T(\y - A\hx)\|_\infty\nn\\
 &\leq & \|A^T\w\|_\infty + \|A^T(\y - A\hx)\|_\infty\nn\\
 & = & \kappa \lambda_n\sigma + \|A^T(\y - A\hx)\|_\infty.
\end{eqnarray}

We again follow the procedure in \cite{candes2009lowrank} (see also \cite{bickel2009simultaneous}) to estimate $\|A^T(\y - A\hx)\|_2$. Since $\hx$ is the solution to \eqref{lasso}, the optimality condition yields that
\begin{eqnarray}
A^T(\y - A\hx) \in \lambda_n\sigma \partial \|\hx\|_1,
\end{eqnarray}
where $\partial \|\hx\|_1 = [-1,1]^n$ is the subgradient of $\|\cdot\|_1$ evaluated at $\hx$.

As a consequence, we obtain
\begin{eqnarray}\label{lassoastarerro}
\|A^T(\y - A\hx)\|_\infty &\leq& \lambda_n\sigma.
\end{eqnarray}

Following the same lines in \eqref{ds_Ah_ubd}, we get
\begin{eqnarray}
\|A\h\|_2^2 \leq (\kappa + 1)\lambda_n\sigma\|\h\|_1.
\end{eqnarray}
Then, Equation \eqref{lassoh1h2}, \eqref{lassoastara} and \eqref{lassoastarerro}
\begin{eqnarray}
  &&\rho_{\frac{4k}{(1-\kappa)^2}}^2\|\h\|_2^2 \leq  \|A\h\|_2^2\nn\\
& \leq & (\kappa + 1)\lambda_n\sigma \frac{\sqrt{4k}}{1-\kappa} \|\h\|_2.
\end{eqnarray}
As a consequence, we get
\begin{eqnarray}
\|\hx - \x\|_2 \leq \frac{1 + \kappa}{1-\kappa}\cdot \frac{2\sqrt{k}}{\rho_{\frac{4k}{(1-\kappa)^2}}^2}\lambda_n\sigma.
\end{eqnarray}
\end{proof}

\bibliographystyle{IEEEbib}
\bibliography{IEEEabrv,Gongbib}

\begin{thebibliography}{10}

\bibitem{Candes2006Uncertainty}
E.~J. Cand\`{e}s, J.~Romberg, and T.~Tao,
\newblock ``Robust uncertainty principles: {E}xact signal reconstruction from
  highly incomplete frequency information,''
\newblock {\em {IEEE} Trans. Inf. Theory}, vol. 52, no. 2, pp. 489--509, Feb.
  2006.

\bibitem{Donoho2006Compressed}
D.~L. Donoho,
\newblock ``Compressed sensing,''
\newblock {\em {IEEE} Trans. Inf. Theory}, vol. 52, no. 4, pp. 1289--1306, Apr.
  2006.

\bibitem{Candes2005Decoding}
E.~J. Cand\`{e}s and T.~Tao,
\newblock ``Decoding by linear programming,''
\newblock {\em {IEEE} Trans. Inf. Theory}, vol. 51, no. 12, pp. 4203--4215,
  Dec. 2005.

\bibitem{Larsson2007Regression}
E.~G. Larsson and Y.~Selen,
\newblock ``Linear regression with a sparse parameter vector,''
\newblock {\em {IEEE} Trans. Signal Process.}, vol. 55, no. 2, pp. 451--460,
  Feb. 2007.

\bibitem{Willsky2005Source}
D.~Malioutov, M.~Cetin, and A.~S. Willsky,
\newblock ``A sparse signal reconstruction perspective for source localization
  with sensor arrays,''
\newblock {\em {IEEE} Trans. Signal Process.}, vol. 53, no. 8, pp. 3010--3022,
  Aug. 2005.

\bibitem{Model2006Signal}
D.~Model and M.~Zibulevsky,
\newblock ``Signal reconstruction in sensor arrays using sparse
  representations,''
\newblock {\em Signal Processing}, vol. 86, pp. 624--638, Mar. 2006.

\bibitem{donoho1989uncertainty}
D.~L. Donoho and P.~B. Stark,
\newblock ``Uncertainty principles and signal recovery,''
\newblock {\em SIAM Journal on Applied Mathematics}, vol. 49, no. 3, pp.
  906--931, 1989.

\bibitem{Baraniuk2007DNA}
M.~A. Sheikh, S.~Sarvotham, O.~Milenkovic, and R.~G. Baraniuk,
\newblock ``{DNA} array decoding from nonlinear measurements by belief
  propagation,''
\newblock in {\em Proc. IEEE Workshop Statistical Signal Processing (SSP
  2007)}, Madison, WI, Aug. 2007, pp. 215--219.

\bibitem{Vikalo2007DNA}
H.~Vikalo, F.~Parvaresh, and B.~Hassibi,
\newblock ``On recovery of sparse signals in compressed {DNA} microarrays,''
\newblock in {\em Proc. Asilomar Conf. Signals, Systems and Computers (ACSSC
  2007)}, Pacific Grove, CA, Nov. 2007, pp. 693--697.

\bibitem{Parvaresh2008DNA}
F.~Parvaresh, H.~Vikalo, S.~Misra, and B.~Hassibi,
\newblock ``Recovering sparse signals using sparse measurement matrices in
  compressed {DNA} microarrays,''
\newblock {\em {IEEE} J. Sel. Topics Signal Processing}, vol. 2, no. 3, pp.
  275--285, June 2008.

\bibitem{Vikalo2008DNA}
H.~Vikalo, F.~Parvaresh, S.~Misra, and B.~Hassibi,
\newblock ``Sparse measurements, compressed sampling, and {DNA} microarrays,''
\newblock in {\em Proc. IEEE Int. Conf. Acoustics, Speech and Signal Processing
  (ICASSP 2008)}, Las Vegas, NV, Apr. 2008, pp. 581--584.

\bibitem{Baraniuk2007Radar}
R.~Baraniuk and P.~Steeghs,
\newblock ``Compressive radar imaging,''
\newblock in {\em IEEE Radar Conference}, Apr. 2007, pp. 128--133.

\bibitem{Herman2008Radar}
M.~Herman and T.~Strohmer,
\newblock ``Compressed sensing radar,''
\newblock in {\em IEEE Radar Conference}, May 2008, pp. 1--6.

\bibitem{Herman2008Highradar}
M.~A. Herman and T.~Strohmer,
\newblock ``High-resolution radar via compressed sensing,''
\newblock {\em {IEEE} Trans. Signal Process.}, vol. 57, no. 6, pp. 2275--2284,
  June 2009.

\bibitem{Tian2007Cognitive}
Z.~Tian and G.~B. Giannakis,
\newblock ``Compressed sensing for wideband cognitive radios,''
\newblock in {\em Proc. IEEE Int. Conf. Acoustics, Speech and Signal Processing
  (ICASSP 2007)}, Honolulu, HI, Apr. 2007, pp. IV--1357--IV--1360.

\bibitem{Sen2010ofdm}
S.~Sen and A.~Nehorai,
\newblock ``Sparsity-based multi-target tracking using ofdm radar,''
\newblock Submitted to \emph{{IEEE Trans. Signal Process.}}

\bibitem{Candes2008RIP}
E.~J. Cand\`{e}s,
\newblock ``The restricted isometry property and its implications for
  compressed sensing,''
\newblock {\em Compte {Rendus} de l'Academie des {Sciences}, {Paris}, {Serie}
  {I}}, vol. 346, 2008.

\bibitem{Cohen2009NSP}
A.~Cohen, W.~Dahmen, and R.~DeVore,
\newblock ``Compressed sensing and best $k$-term approximation,''
\newblock {\em J. Amer. Math. Soc.}, vol. 22, pp. 211--231, July 2009.

\bibitem{dAspremont2007sparsePCA}
A.~d'Aspremont, L.~El Ghaoui, M.~Jordan, and G.~R.~G. Lanckriet,
\newblock ``A direct formulation for sparse {PCA} using semidefinite
  programming,''
\newblock {\em SIAM Review}, vol. 49, no. 3, pp. 434--448, 2007.

\bibitem{dAspermont2010Nullspace}
A.~{d'Aspremont} and L.~{El Ghaoui},
\newblock ``Testing the nullspace property using semidefinite programming,''
\newblock {\em ArXiv e-prints}, Nov. 2010.

\bibitem{Juditsky2010Verifiable}
A.~{Juditsky} and A.~S. {Nemirovski},
\newblock ``On verifiable sufficient conditions for sparse signal recovery via
  $\ell_1$ minimization,''
\newblock {\em ArXiv e-prints}, May 2010.

\bibitem{Donoho1998Atomic}
S.~Chen, D.~L. Donoho, and M.~A. Saunders,
\newblock ``Atomic decomposition by basis pursuit,''
\newblock {\em {SIAM} J. Sci. Comp.}, vol. 20, no. 1, pp. 33--61, 1998.

\bibitem{Candes2007Dantzig}
E.~J. Cand\`{e}s and T.~Tao,
\newblock ``The {Dantzig} selector: Statistical estimation when $p$ is much
  larger than $n$,''
\newblock {\em Ann. Statist.}, vol. 35, pp. 2313--2351, 2007.

\bibitem{Tibshirani1996Lasso}
R.~Tibshirani,
\newblock ``Regression shrinkage and selection via lasso,''
\newblock {\em J. Roy. Statist. Soc. Ser. B}, vol. 58, pp. 267--288.

\bibitem{bickel2009simultaneous}
P.~Bickel, Y.~Ritov, and A.~Tsybakov,
\newblock ``Simultaneous analysis of {Lasso} and {Dantzig} selector,''
\newblock {\em Annals of Statistics}, vol. 37, no. 4, pp. 1705--1732, 2009.

\bibitem{bickel2007discussion}
P.~J. Bickel,
\newblock ``Discussion of the {Dantzig} selector: statistical estimation when
  $p$ is much larger than $n$, by e. j. cand\`es and t. tao,''
\newblock {\em Annals of Stat.}, pp. 2352--2357, 2007.

\bibitem{meinshausen2009lassotype}
N.~Meinshausen and B.~Yu,
\newblock ``Lasso-type recovery of sparse representations for high-dimensional
  data,''
\newblock {\em Ann. Statist.}, vol. 37, pp. 246--270, 2009.

\bibitem{Baraniuk2008RIP}
R.~DeVore R.~Baraniuk, M.~Davenport and M.~B. Wakin,
\newblock ``A simple proof of the restricted isometry property for random
  matrices,''
\newblock {\em Constructive Approximation}, vol. 28, no. 3, pp. 253--263, 2008.

\bibitem{mendelson2007subgaussian}
S.~Mendelson, A.~Pajor, and N.~Tomczak-Jaegermann,
\newblock ``Reconstruction and subgaussian operators in asymptotic geometric
  analysis,''
\newblock {\em Geometric And Functional Analysis}, pp. 1248--1282, Nov. 2007.

\bibitem{zhang2005overunder}
Y.~Zhang,
\newblock ``A simple proof for recoverability of $\ell_1$-minimization: go over
  or under?,''
\newblock Tech. {R}ep., Rice CAAM Department, 2005.

\bibitem{donoho2001uncertainty}
D.~L. Donoho and X.~Huo,
\newblock ``Uncertainty principles and ideal atomic decomposition,''
\newblock {\em {IEEE} Trans. Inf. Theory}, vol. 47, no. 7, pp. 2845--2862, Nov.
  2001.

\bibitem{donoho2004highdimensional}
D.~Donoho,
\newblock ``High-dimensional centrally-symmetric polytopes with neighborliness
  proportional to dimension,''
\newblock Technical report, Department of Statistics, Stanford University,
  2004.

\bibitem{mendelson2008embedding}
S.~Mendelson and N.~Tomczak-Jaegermann,
\newblock ``A subgaussian embedding theorem,''
\newblock {\em Israel Journal of Mathematics}, pp. 349--364, Mar. 2008.

\bibitem{talagrand2005generic}
M.~Talagrand,
\newblock {\em The generic chaining: upper and lower bounds of stochastic
  processes},
\newblock Springer, 2005.

\bibitem{fernique1975regularite}
X.~Fernique,
\newblock {\em R\'{e}gularit\'{e} des trajectoires des fonctiones
  al\'{e}atoires gaussiennes},
\newblock Ecole d'Et\'{e} de {Probabilit\'{e}s} de {St}-Flour 1974, {Lecture}
  {Notes} in {Mathematics} 480, {Springer}-Verlag, 1975.

\bibitem{talagrand1987regularity}
M.~Talagrand,
\newblock ``Regularity of {Gaussian} processes,''
\newblock {\em Acta Math.}, vol. 159, pp. 99--149, 1987.

\bibitem{Bodlaender1990Normmaximization}
H.~Bodlaender, P.~Gritzmann, V.~Klee, and J.~Leeuwen,
\newblock ``Computational complexity of norm-maximization,''
\newblock {\em Combinatorica}, vol. 10, pp. 203--225, 1990.

\bibitem{boyd2004convex}
S.~Boyd and L.~Vandenberghe,
\newblock {\em Convex Optimization},
\newblock Cambridge University Press, 2004.

\bibitem{Nash1996Programming}
A.~Sofer S.~G.~Nash,
\newblock {\em Linear and nonlinear programming},
\newblock McGraw-Hill, New York, NY, 1996.

\bibitem{byrd2000a}
R.~H. Byrd, J.~C. Gilbert, and J.~Nocedal,
\newblock ``A trust region method based on interior point techniques for
  nonlinear programming,''
\newblock {\em Mathematical Programming}, vol. 89, no. 1, pp. 149--185, 2000.

\bibitem{byrd1999an}
R.~H. Byrd, Mary~E. Hribar, and Jorge Nocedal,
\newblock ``An interior point algorithm for large-scale nonlinear
  programming,''
\newblock {\em SIAM Journal on Optimization}, vol. 9, no. 4, pp. 877--900,
  1999.

\bibitem{waltz2006an}
R.~A. Waltz, J.~L. Morales, J.~Nocedal, and D.~Orban,
\newblock ``An interior algorithm for nonlinear optimization that combines line
  search and trust region steps,''
\newblock {\em Mathematical Programming}, vol. 107, no. 3, pp. 391--408, 2006.

\bibitem{schrijver1991cones}
L.~Lov\'asz and A.~Schrijver,
\newblock ``Cones of matrices and set-functions and 0-1 optimization,''
\newblock {\em SIAM Journal on Optimization}, vol. 1, pp. 166--190, 1991.

\bibitem{alizadeh1995interior}
F.~Alizadeh,
\newblock ``Interior point methods in semidefinite programming with
  applications to combinatorial optimization,''
\newblock {\em SIAM Journal on Optimization}, vol. 5, pp. 13--51, 1995.

\bibitem{oustry1999semidefinite}
C.~Lemar\'echal and F.~Oustry,
\newblock ``Semidefinite relaxations and {Lagrangian} duality with application
  to combinatorial optimization,''
\newblock {\em INRIA, {Rapport} de recherche}, vol. 3710, 1999.

\bibitem{Sturm1999Sedumi}
J.~Sturm,
\newblock ``Using {SEDUMI} 1.0x, a {MATLAB} toolbox for optimization over
  symmetric cones,''
\newblock {\em Optimization Methods and Software}, vol. 11, pp. 625--653, 1999.

\bibitem{Tutuncu2003SDPT3}
R.~H Tutuncu, K.~C. Toh, and M.~J. Todd,
\newblock ``Solving semidefinite-quadratic-linear programs using {SDPT}3,''
\newblock {\em Mathematical Programming Ser. B}, vol. 95, pp. 189--217, 2003.

\bibitem{Grant2009CVX}
M.~Grant and S.~Boyd,
\newblock ``{CVX}: {Matlab} software for disciplined convex programming (web
  page and software),''
\newblock June 2009.

\bibitem{Andersenmosek}
K.~Andersen E.~Andersen,
\newblock ``The {MOSEK} optimization tools mannual, {Version 5.0},''
\newblock http://www.mosek.com.

\bibitem{candes2009lowrank}
E.~J. Cand\`es and Y.~Plan,
\newblock ``Tight oracle bounds for low-rank matrix recovery from a minimal
  number of random measurements,''
\newblock Submitted for publication, 2009.

\end{thebibliography}
\end{document}